\documentclass{article}
\usepackage{arxiv}

\usepackage[utf8]{inputenc} % allow utf-8 input
\usepackage[T1]{fontenc}    % use 8-bit T1 fonts
\usepackage{hyperref}       % hyperlinks
\usepackage{url}            % simple URL typesetting
\usepackage{booktabs}       % professional-quality tables
\usepackage{amsfonts}       % blackboard math symbols
\usepackage{nicefrac}       % compact symbols for 1/2, etc.
\usepackage{microtype}      % microtypography
\usepackage{lipsum}
\usepackage{graphicx}
\graphicspath{ {./images/} }

\title{Anisotropic--Isotropic Transition of Cages at the Glass Transition}

\author{
 Huijun Zhang \\
  State Key Laboratory for Mechanical Behavior of Materials\\
  Shaanxi International Research Center for Soft Matter\\
  School of Materials Science and Engineering\\
  Xi'an Jiaotong University\\
  Xi'an, 710049, China\\
  \texttt{huijun@xjtu.edu.cn} \\
   \And
 Feng Liu \\
  State Key Laboratory for Mechanical Behavior of Materials\\
  Shaanxi International Research Center for Soft Matter\\
  School of Materials Science and Engineering\\
  Xi'an Jiaotong University\\
  Xi'an, 710049, China\\
  \texttt{feng.liu@xjtu.edu.cn} \\
  \And
 Yilong Han \\
  Department of Physics\\
  Hong Kong University of Science and Technology\\
  Clear Water Bay\\
  Kowloon\\
  Hong Kong\\
  \texttt{yilong@ust.hk} \\
}

\begin{document}
\maketitle
\begin{abstract}
Characterizing the local structural evolution is an essential step in understanding the nature of glass transition. In this work, we probe the evolution of Voronoi cell geometry in simple glass models, and find that the individual particle cages deform anisotropically in supercooled liquid and isotropically in glass. We introduce an anisotropy parameter $k$ for each Voronoi cell, which mean value exhibits a sharp change at the mode-coupling glass transition $\phi_\mathrm{c}$. Moreover, a power law of packing fraction $\phi\propto q_1^{-d}$ is discovered in the liquid regime with $d>D$, in contrast to $d=D$ in the glass regime, where $q_1$ is the first peak position of structure factor, and $D$ is the space dimension. This power law is explained by the change of $k$. The active motions in supercooled liquid are spatially correlated with long axes rather than short axes of Voronoi cells. In addition, the dynamic slowing down approaching the glass transition can be well characterized through a modified free volume model based on $k$. These findings reveal that the nonagnostic structural parameter $k$ determines glassy dynamics and is effective in identifying the structure-dynamics correlations and the glass transition.  
\end{abstract}

% keywords can be removed
%\keywords{First keyword \and Second keyword \and More}

\section{Introduction}
Glasses or amorphous solids are produced by cooling \cite{berthier11} or densifying \cite{parisi10RMP} liquids. As the glass transition is approached, dynamics slows down drastically without any noticeable structural change \cite{dyre06,binder11}. The mode-coupling theory (MCT) \cite{gotze09, luo22} or Vogel-Fulcher-Tamman law \cite{berthier11} predicts the drastic growth of relaxation time $\tau_\mathrm{\alpha}$ before the glass transition but do not provide structural features for the slowing down. Although experiments \cite{hecksher08,brambilla09,tanaka10,vivek17,hallett18,tah18} and simulations \cite{kob13,biroli13,yu15,ninarello17,tong18,wang18,marin20,berthier23} have made intensive efforts in searching for effective structural parameters in characterizing the glass transition, the structural mechanism of glass transition remains unclear. Whether glass transition has a structural origin or is purely dynamic remains controversial \cite{berthier11, royall15}.

Various parameters, such as two-point entropy \cite{tanaka10,zheng21}, soft spot from vibration modes \cite{widmer08}, and neighboring structures from machine learning \cite{schoenholz16,bapst20}, have been proposed to link local structures with dynamic slowing down, but these approaches are structural agnostic \cite{tanaka10,zheng21} and lack intuitive structural features. Great efforts have been devoted to identify nonagnostic locally favored structures (LFSs) responsible for the dynamic slowing down. For instance, icosahedral \cite{frank52,royall15}, crystalline \cite{tanaka10,tah18}, and tetrahedral clusters \cite{xia15, marin20} have been found to correlate with the slow dynamics in different glass-forming liquids, but these structural signatures are system-dependent \cite{hocky14} and cannot distinguish glass structure from liquid because these motifs similarly exist in both liquid and glass regimes. In addition, structural indicators of icosahedra or tetrahedra \cite{frank52,royall15,xia15,marin20} are not applicable for 2D systems. Recently, a quantitative relation between structural order and slow glassy dynamics for both 2D and 3D systems has been constructed by measuring particle’s deviation from the reference cluster \cite{tong18}, but it is also order-agnostic without a prior knowledge of the preferred local structure \cite{tanaka19}. Therefore, a structure parameter that can reveal the local structural changes in glass transition without the above limitations is needed. 

Another topic in glass studies is the structural power law \cite{wang09,zeng16}. For a $D$-dimensional crystal, its mean atomic volume must satisfy the power law, $v_\mathrm{a}\propto a^D\propto q_1^{-D}\propto \phi^{-1}$, where $a$ is the lattice constant. $q_1$ is the first peak position of the structure factor $S(q)$, and $\phi$ is the packing fraction. Amorphous solids also follow the power law \cite{wang09},
\begin{equation}\label{nsl}
{q_1}\propto \phi^\frac{1}{d}.
\end{equation}
However, $d\ne D$ in many glasses \cite{wang09,zeng14,chen15,zeng16,xia17,zhang20,zhang23}, which gives rise to controversies on its mechanism. Recent studies showed that pressure (or equivalently, density) and composition changes lead to different power laws. Under pressure change, $d=D$ in glasses for particles with the same softness, but deviates from $D$ when the system is composed of hard and soft particles \cite{zhang20}. Link $q_1$ in the reciprocal space to certain structural change in real space is difficult because $S(q_1)$ contains structures spanning broad length scales in real space. How the local deformations affect the power law in reciprocal space remains unclear. Moreover, whether the power law holds in liquid regime and how it changes at the glass transition have not been explored. 

\begin{figure}[t] 
\centering
\includegraphics[width=.6\columnwidth]{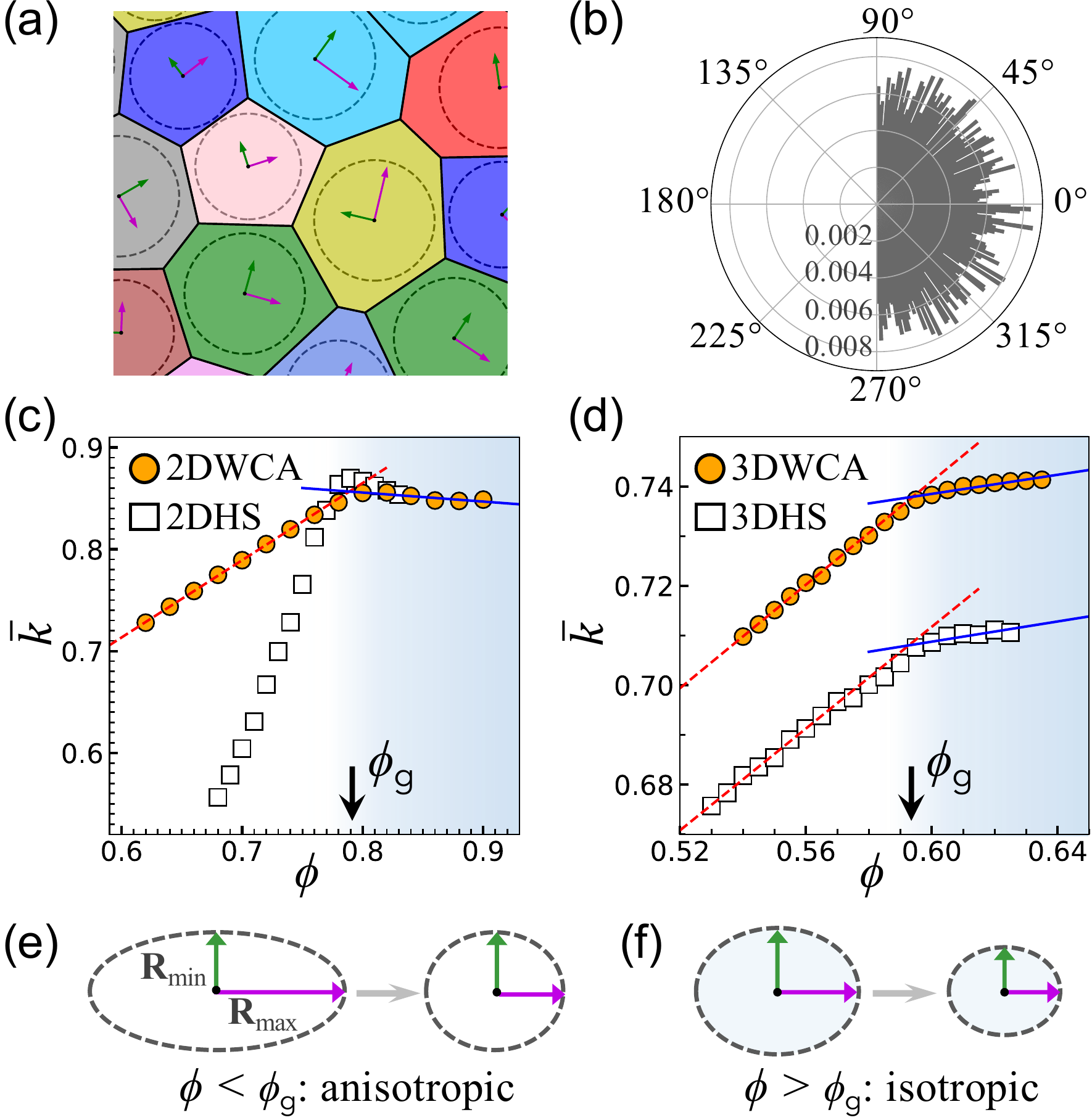} 
\caption{Anisotropic-to-isotropic compression across the glass transition. (a) Radical Voronoi polygons of a subarea of the 2DWCA liquid at $\phi=0.74$. The large eigenvector $\bf{R}_\mathrm{max}$ (pink arrow) and small eigenvector $\bf{R}_\mathrm{min}$ (green arrow) of the Voronoi polygons are labeled at the center of each particle (dashed circle). (b) The angular distribution of $\mathbf{R}_\mathrm{max}$. $\bf{R}_\mathrm{min}$ is always perpendicular to $\mathbf{R}_\mathrm{max}$. (c) Average anisotropy $\bar{k}$ as a function of $\phi$ for 2DWCA (orange circles) and 2DHS (open squares) systems with size ratio $\lambda=1.3$ and fraction of large particles $x=0.5$. The liquid (red line) and glass (blue line) behavior of 2D glasses intersects at $\phi_\mathrm{g}\approx0.79$. (d) Average anisotropy $\bar{k}$ as a function of $\phi$ for 3DWCA (orange circles) and 3DHS (open squares) systems with $\lambda=1.3$ and $x=0.5$. The liquid (red line) and glass (blue line) regimes for 3D systems intersect at $\phi_\mathrm{g}\approx0.59$. (e) and (f) Schematic of the anisotropic deformation at $\phi<\phi_\mathrm{g}$ and isotropic deformation at $\phi>\phi_\mathrm{g}$, respectively.}
\label{fig1}
\end{figure}

Each particle in a supercooled liquid is caged by its nearest neighboring particles \cite{mayer08}. The cage effect has been widely used to qualitatively explain the dynamical behaviors \cite{weeks00,chaudhuri15} and plastic deformations \cite{laurati17,yu18}. However, a direct link between cage structure and glass transition remains unavailable. The cage of each particle is described by its Voronoi cell \cite{voro} [Fig.~\ref{fig1}(a)], which reflects the particle’s free volume not shared by its neighbors. Voronoi tessellation is usually applied to characterize the local packing of particles, which has been used to understand the jamming transition \cite{rieser16} and Boson peak \cite{yang19} in amorphous solids. Here, we find that the cage anisotropy can effectively reveal the structure change at the glass transition.

\section{Results}
\paragraph{Methods}
We perform molecular dynamics simulations \cite{plimpton95} for binary Weeks–Chandler–Andersen (WCA) particles \cite{weeks71} and event-driven molecular dynamics simulations \cite{bannerman11} for binary hard spheres (HSs) in both two (2D) and three dimensions (3D). Binary sized particles are used to avoid crystallization. Systems are investigated in the $NVT$ ensemble under periodic boundary conditions. 2D and 3D glassy systems have similar behavior and caging effects \cite{shiba16,vivek17,illing17}. Hence, we focus on 2DWCA systems as examples with various diameter ratios $\lambda$ and fractions of large spheres $x$. The simulation details are provided in the Supplemental Material. 

\subsection{Cage anisotropy parameter $k$}
We use radical Voronoi tessellation \cite{voro} [Fig.~\ref{fig1}(a)] rather than traditional Voronoi tessellation for binary systems to avoid the incorrect bisecting line cutting through a large sphere. The asymmetric shape of the Voronoi cell is characterized by a tensor \cite{mayer08,schroder10}, $\bf{I}=\int_V \mathbf{r}\otimes\mathbf{r}~\mathbf{d}r$, which integrates over the cell volume $V$; $\mathbf{r}$ is the position vector relative to the particle’s center, and $\otimes$ denotes a dyadic product. $\bf{I}$ represents the shape of the Voronoi cell, and its principal axes are obtained by diagonalizing $\bf{I}$. The two orthogonal eigenvectors, $\mathbf{R}_\mathrm{min}$ and $\mathbf{R}_\mathrm{max}$, represent the long and short axes of the cage, respectively [Fig.~\ref{fig1}(a)]. Their ratio defines the cage anisotropy parameter 
\begin{equation}
k\equiv\frac{|\bf{R}_\mathrm{min}|}{|\bf{R}_\mathrm{max}|}\in (0,1].
\end{equation}
$k=1$ for an isotropic cage, and the deviation from 1 quantifies the elongation anisotropy of the cage. Similar parameters about the asymmetric elongation of the cage have been used to study soft colloidal glasses \cite{mayer08} and jamming transition in 3D HS systems \cite{schroder10}. However, the causal link between the anisotropy parameter and glass transition has rarely been explored. Figure~\ref{fig1}(a) shows that $\mathbf{R}_\mathrm{min}$ points to its neighboring particles, whereas $\mathbf{R}_\mathrm{max}$ points to the gap between two neighbors. The uniform distribution of $\mathbf{R}_\mathrm{max}$ directions in Figs.~\ref{fig1}(a) and (b) manifests the uniformity of the liquid. 

\subsection{Anisotropy and volume of cages at glass transition}
As $\phi$ increases, free volumes around particles are squeezed, yielding more isotropic cages, that is, higher averaged $\bar{k}$. In Fig.~\ref{fig1}(c), $\bar{k}(\phi)$ increases and reaches a plateau  at $\phi_ k> 0.79$ for 2DWCA and 2DHS systems, demonstrating a structure change at their glass transitions. The increase in $\bar{k}(\phi)$ at $\phi<\phi_\mathrm{g}$ in Fig.~\ref{fig1}(c) shows that cages are compressed more along $\mathbf{R}_\mathrm{max}$ than $\mathbf{R}_\mathrm{min}$ direction, that is, the cage deformation is anisotropic, as illustrated in Fig.~\ref{fig1}(e). In the liquid regime, $\bar{k}$ is higher in 2DWCA than that in 2DHS system [Fig.~\ref{fig1}(c)] because the softer WCA particles can adjust their positions more easily to form more homogenous structures with more isotropic cages. At $\phi>\phi_\textrm{g}$, $\bar{k}$ is comparable for 2DWCA and 2DHS systems because both WCA and HS particles interact via their hard cores at high $\phi$. $\bar{k}$ is approximately constant at $\phi>\phi_\mathrm{g}$ [Fig.~\ref{fig1}(c)], indicating that cages isotropically deform in glasses [Fig.~\ref{fig1}(f)]. Similar behavior of $\bar{k}$ is also observed in 3DWCA and 3DHS systems [Fig.~\ref{fig1}(d)] and 2DWCA systems with different $\lambda$ (Supplemental Fig.~S1). $\bar{k}$ is lower in 3D system than that in the 2D systems [Figs.~\ref{fig1}(c) and \ref{fig1}(d)] because particles in the 3D system are more disordered and more difficult to adjust their positions due to stronger confinement from more neighboring particles. 

\begin{figure}[!t] 
\centering
\includegraphics[width=0.6\columnwidth]{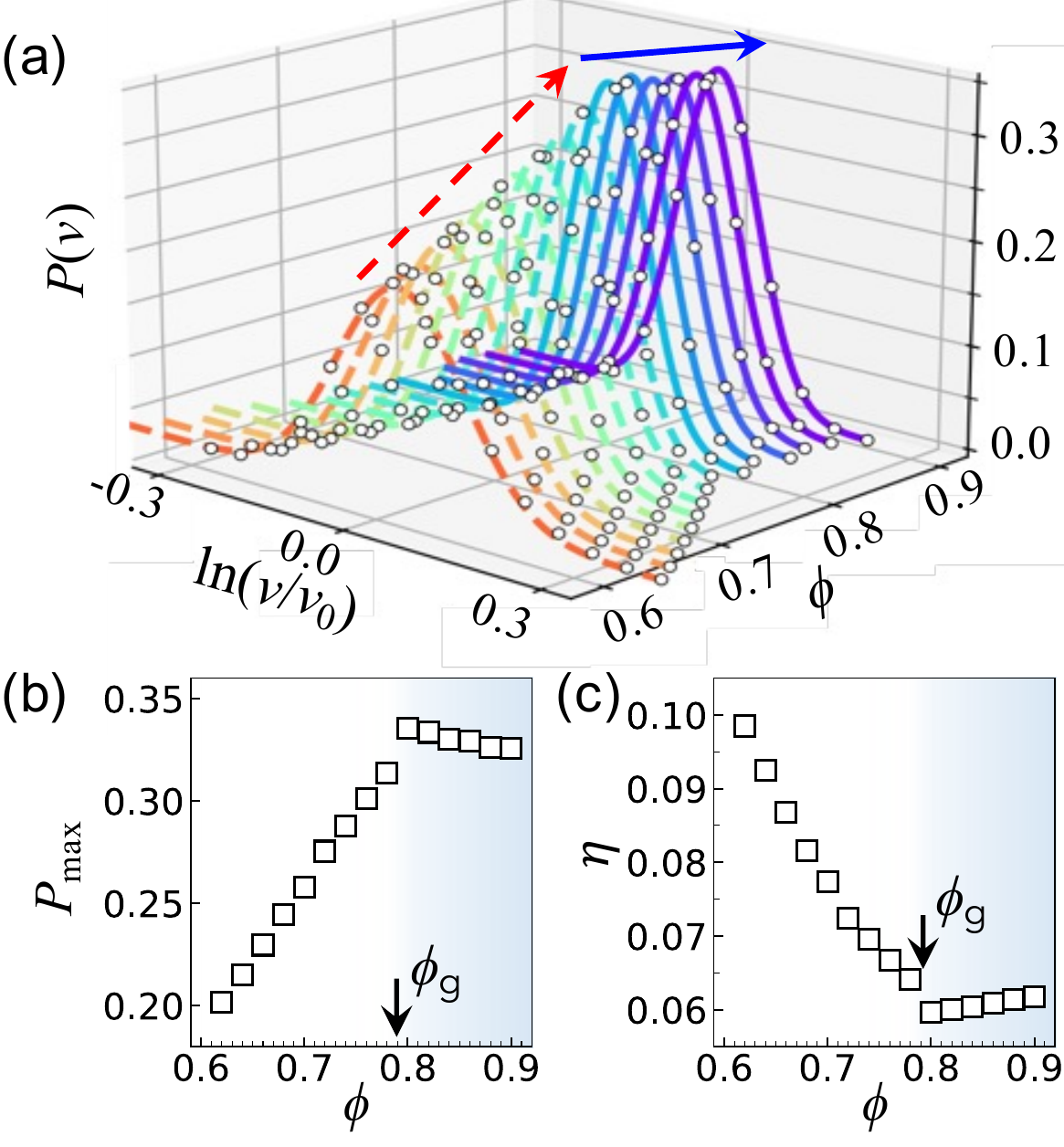} 
\caption{Voronoi cell volume $v$ in 2DWCA systems. (a) Distributions of $v$ fitted by Eq.~(\ref{distr}) (curves). $P(v)$ barely changes with $\phi$ at $\phi> 0.79$. Red and blue arrows show the increasing trend of the peaks in anisotropic ($\phi<0.79$) and isotropic ($\phi>0.79$) regimes, respectively. (b) The fitted maximum of $P(v)$. (c) The fitted standard deviation of $P(v)$.} 
\label{fig2}
\end{figure}

The anisotropic-to-isotropic transition of cage deformation [Figs.~\ref{fig1}(c) and \ref{fig1}(d)] is further verified by the Voronoi cell volume $v$.  Figure~\ref{fig2}(a) shows that $v$ of large particles satisfies a log-normal distribution well at each $\phi$, 
\begin{equation}\label{distr}
P(v)=\frac{1}{\sqrt{2\pi}\eta}\mathrm{exp}(-\frac{(\mathrm{ln}v-\mathrm{ln}v_0)^2}{2\eta^2}).
\end{equation}
$P(v)$ barely changes at $\phi>0.79$ in Fig.~\ref{fig2}(a), implying isotropic shrinkages of cages, in accordance with Fig.~\ref{fig1}(c). By contrast, the anisotropic deformations of cages at $\phi<0.79$ change $P(v)$, as shown in Fig.~\ref{fig2}(a). The height and standard deviation of $P(v)$ exhibit a clear change at $\phi_\mathrm{g}=0.79$ [Figs.~\ref{fig2}(b) and \ref{fig2}(c)], which is consistent with Fig.~\ref{fig1}(c). $P(v)$ of small particles shows the same transition at $\phi_\mathrm{g}$ (Supplemental Fig.~S2). The previous work scaled the distribution along the $v$ axis with the standard deviation \cite{Starr02} and did not report the transition shown in Fig.~\ref{fig2}. 

\begin{figure}[!b] 
\centering
\includegraphics[width=0.7\columnwidth]{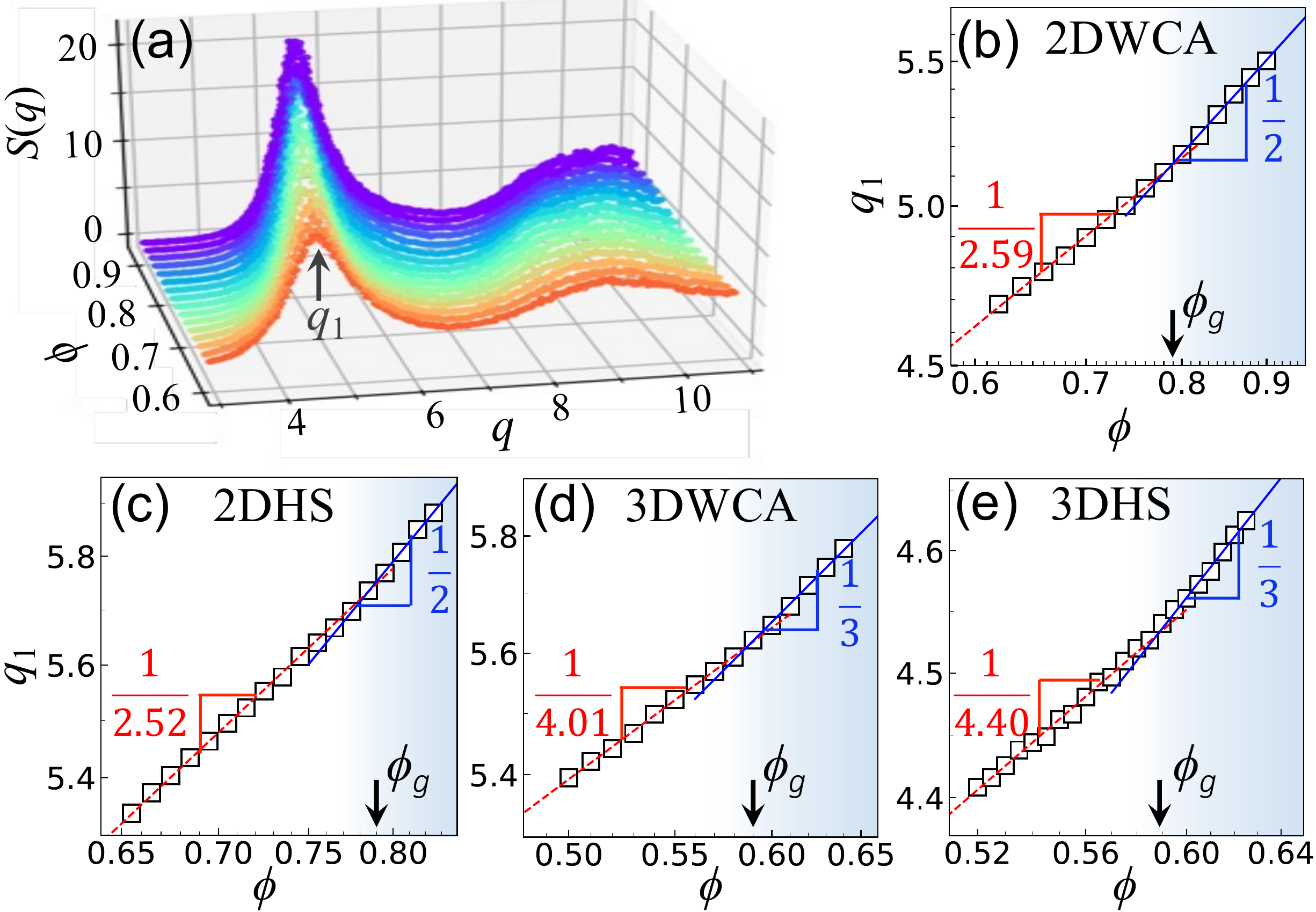}
\caption{Structural power law of Eq.~(\ref{nsl}). (a) $S(q)$ at $0.62\le\phi\le 0.92$ for 2DWCA systems. (b)–(e) Log–log plots of $q_1(\phi)$ fitted with Eq.~(\ref{nsl}) (red and blue lines) for 2DWCA, 2DHS, 3DWCA, and 3DHS systems. The blue line with $d=D$ and the red line with $d>D$ intersect at $\phi_\mathrm{g}$. The values of $1/d$, that is, the slopes, are labeled in the figures.} 
\label{fig3}
\end{figure}

\subsection{Structural power law}
We generalize the power law Eq.~(\ref{nsl}) from glass \cite{zeng14,xia17,zhang20} to liquid. $q_1$ in Eq.~(\ref{nsl}) is measured from the structure factor $S(q)=\langle \sum_{j=1}^{N}e^{i\mathbf{q}\cdot \mathbf{r}_j}\sum_{k=1}^{N}e^{-i\mathbf{q}\cdot \mathbf{r}_k} \rangle/N$. As $\phi$ increases, the interparticle distance decreases, and $q_1$ increases [Fig.~\ref{fig3}(a)]. For glasses composed of binary particles with the same softness, the compression is locally uniform, and thus $d=D$ \cite{zhang20}. Large and small spheres in 2DWCA systems have the same WCA potential, that is, the same softness, so the compression-induced deformation is spatially uniform \cite{zhang20}. Consequently, the exponent $d=D=2$, which is confirmed in the glass regime at $\phi_\mathrm{g}> 0.79$ [Fig.~\ref{fig3}(b)]. The power law still holds in the liquid regime, however, $d=2.59>D$ for 2DWCA. Such $d=D$ in the glass regime and $d>D$ in the liquid regime have also been observed in 2DHS [Fig.~\ref{fig3}(c)], 3DWCA [Fig.~\ref{fig3}(d)], 3DHS [Fig.~\ref{fig3}(e)], and 2DWCA systems at different $\lambda$ and $x$ (Supplemental Fig.~S3). Thus, the coincidence between the glass transition and anisotropic–isotropic transition of cage deformation is robust. 

$d$ deviating from $D$ has been observed in various metallic glasses under composition change \cite{wang09,zhang23} or glasses composed of particles with different softness under pressure change \cite{zeng14,zhang20}. Such deviation manifests nonuniform deformations, but the type of deformation that can increase or decrease $d$ is unclear \cite{zeng14,chen15}. Here, we find that $d>D$ in liquid [Figs.~\ref{fig3}(b)–\ref{fig3}(e)] arises from local anisotropic deformation [Fig.~\ref{fig1}(e)] as follows. For a global density change, the compressed volume $\Delta v$ contributed from one anisotropic cage is equivalent to those from several isotropic cages because anisotropic cages usually have more free volumes than isotropic cages. Furthermore, a large $\Delta v$ concentrated at one place (i.e., one anisotropic cage) affects less on $q_1$ than multiple small $\Delta v$ spread in the sample (i.e., multiple isotropic cages). Therefore, the more anisotropic cages in liquid lower the slope of $q_1(\phi)$, that is, increase $d$ [Figs.~\ref{fig3}(b)–\ref{fig3}(e)]. Consequently, the transition between $d>D$ and $d=D$ [Figs.~\ref{fig3}(b)–\ref{fig3}(e)] corresponds to the anisotropic–isotropic transition of cages at the glass transition. Eq.~(\ref{nsl}) arises from the structure change rather than the structure at a certain $\phi$; thus, it is not related to any fractal structure in glass and $d$ can be greater than $D$ \cite{zhang23}.

\begin{figure}[!b] 
\centering
\includegraphics[width=.6\columnwidth]{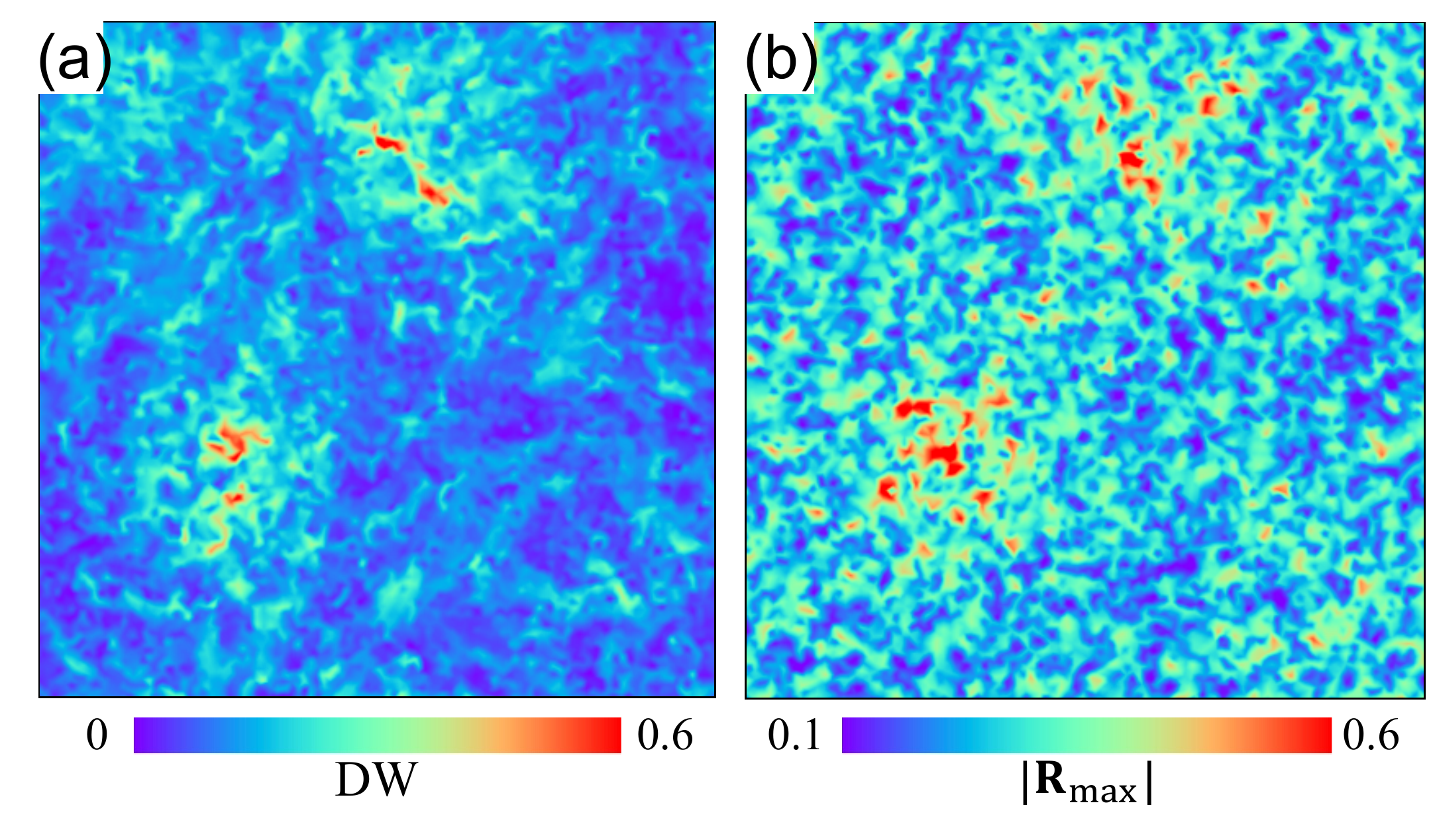} 
\caption{Spatial correlation between dynamics ($DW$) and structure ($|\mathbf{R}_\textrm{max}|$) in the 2DWCA system at $\phi=0.74$. (a) Spatial distribution of the DW factor. Red denotes strong dynamics. (b) Spatial distribution of $|\mathbf{R}_\textrm{max}|$.} 
\label{fig4}
\end{figure}

\subsection{Structure-dynamics correlations}
\begin{figure}[!t] 
\centering
\includegraphics[width=0.6\columnwidth]{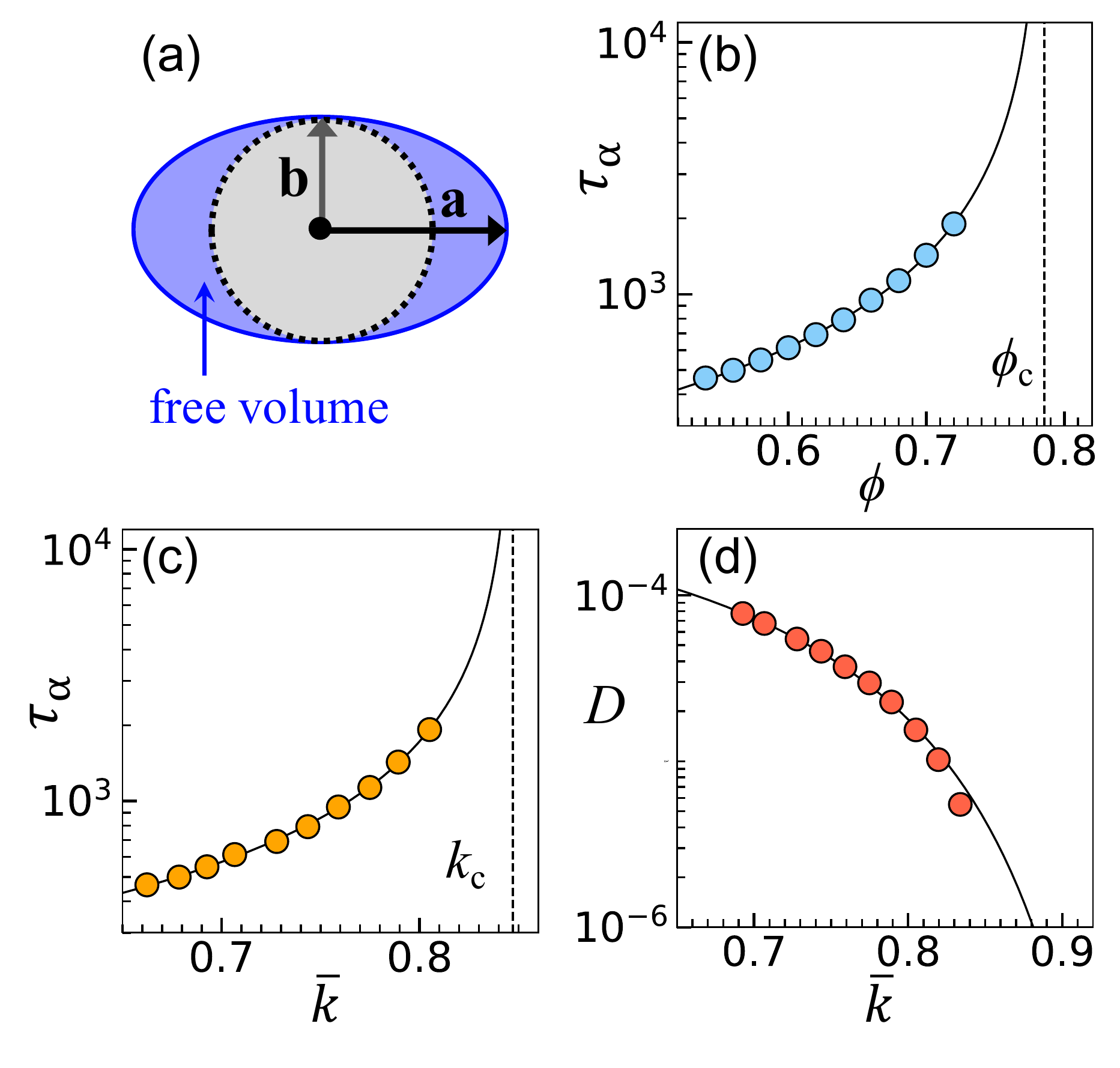} 
\caption{Dynamical behavior of 2DWCA systems. (a) Schematic of the anisotropic cage (ellipse) composed of the particle (gray circle) and free volume (blue region). (b) Relaxation time $\tau_\mathrm{\alpha}(\phi)$ with the MCT fit (curve) yields $\phi_\mathrm{c}=0.785$ and $\gamma=1.07$. (c) $\tau_\mathrm{\alpha}(\bar{k})$ fitted with Eq.~(\ref{mct}) (curve) yields $k_\mathrm{c}=0.847$ and $\gamma=0.97$. (d) Diffusion constant fitted with Eq.~(\ref{D1}) (curve).} 
\label{fig5}
\end{figure}

The dynamics of particle $i$ can be characterized by the Debby–Waller factor $\mathrm{DW}_i = \sqrt{\langle (\mathbf{r} - \langle \mathbf{r} \rangle)^2 \rangle}/\bar{\sigma}$, where $\langle \quad \rangle$ denotes the average over a time before cage breaking (Supplementary Information) \cite{cao17}, and $\bar{\sigma}$ is the average particle diameter. Particles with large DW factors (i.e., strong dynamics) in Fig.~\ref{fig4}(a) are spatially localized and more correlated with the large-$|\mathbf{R}_\mathrm{max}|$ regions in Fig.~\ref{fig4}(b), implying that particles tend to move along $\mathbf{R}_\mathrm{max}$ rather than $\mathbf{R}_\mathrm{min}$ (Supplemental Fig.~S4). Therefore, $\mathbf{R}_\mathrm{max}$ reflects a particle’s dynamical propensity. This is reasonable because more free volume distributes along $\pm\mathbf{R}_\mathrm{max}$ [Fig.~\ref{fig5}(a)].

The structure relaxation time $\tau_\alpha$ is measured as the decay time of the self-intermediate scattering function $F_s(q=q_1,t)=\langle \sum_j \mathrm{exp}(i\mathbf{q}\cdot[\mathbf{r}_j(t) - \mathbf{r}_j(0)])/N \rangle$ [Supplemental Fig.~S5(a)]. $\tau_{\alpha}$ can be fitted by the MCT relation \cite{gotze09,berthier11}
$\tau_\alpha \sim (\phi_\mathrm{c}-\phi)^{-\gamma}$ with $\phi_\mathrm{c}= 0.785$ [Fig.~\ref{fig5}(b)]. It coincides well with the structure transitions in Figs.~\ref{fig1}(c), \ref{fig2}(b), \ref{fig2}(c), and \ref{fig3}(b), that is, $\phi_\mathrm{c} = \phi_\mathrm{g}$. Considering that $\bar{k} (\phi)$ is linear in the supercooled liquids [Fig.~\ref{fig1}(c)], the MCT relation can be rewritten as
\begin{equation}\label{mct} 
\tau_\alpha \sim ( k_\mathrm{c}- \bar{k})^{-\gamma}.
\end{equation}
In contrast to the classic MCT relation based on the global thermodynamic parameter $\phi$ (or $T$), Eq.~(\ref{mct}) links the glassy dynamics to specific local structure at the single particle level in supercooled liquids. The fitted $ k_\mathrm{c}\approx 0.847$ [Fig.~\ref{fig5}(c)] is consistent with the measured $\bar{k}=0.85$ in the glass regime [Fig.~\ref{fig1}(c)]. We attribute the increasing dynamic slowing down in supercooled liquids to the increase in $\bar{k}$, that is, cages becoming more isotropic [Fig.~\ref{fig1}(f)], which reduces the free volume and mobility of particles. 

\subsection{Free volume model}
The free volume mode is one of the most widely used and simple theories for the glass transition \cite{turnbull61,berthier11,wang12}, which attributes the dynamic slowing down to the decreased ‘‘free volume to move’’ for particles as $\phi$ increases \cite{berthier11}. If the free volume is known, then it can predict the average long-time diffusion constant of particles as $D(v_\mathrm{f})=A\cdot\mathrm{exp}(\frac{-B}{v_\mathrm{f}})$, where $v_\mathrm{f}=( \bar{v} - v_\mathrm{0})/ v_\mathrm{0}$ is the reduced free volume. $v_\mathrm{0}$ is the volume of a particle, and $\bar {v} $ is the specific volume \cite{turnbull61}. However, this model does not consider the packing structure \cite{turnbull61}, and $v_\mathrm{f}$ is unavailable. Here, we suggest that $v_\mathrm{f}$ can be evaluated on the basis on $k$ of the local cage. Considering that free volumes mainly locate along the long axis, the reduced free volume can be derived from our ellipse approximation of the cage, as shown in Fig.~\ref{fig5}(a), $v_\mathrm{f}=\frac {\pi \cdot |\bf{a}|\cdot |\bf{b}| - \pi \cdot |\bf{b}|^2}{\pi \cdot |\bf{b}|^2} = \frac{1}{ k}-1$ with $\frac{\bf{R}_\mathrm{min}}{\bf{R}_\mathrm{max}} =\frac{\bf{b}}{\bf{a}}$, that is, the blue region in Fig.~\ref{fig5}(a). This result is robust for 3D spheroid approximation: $v_\mathrm{f}=\frac {\frac{4}{3}\pi \cdot |\bf{a}|\cdot |\bf{b}|^2 - \frac{4}{3}\pi \cdot |\bf{b}|^3}{\frac{4}{3}\pi \cdot |\bf{b}|^3} = \frac{1}{ k}-1$. Substituting $v_\mathrm{f}$ to $D(v_\mathrm{f})$ \cite{turnbull61} gives
\begin{equation}\label{D1}
D(\bar{k})=A\cdot\mathrm{exp}(\frac{-B\bar{k}}{1-\bar{ k}}).
\end{equation}
$D(v_\mathrm{f})$ is mainly used to estimate the free volume $v$ from the measured $D$ \cite{turnbull61} because $v$ is not available directly. By contrast, Eq.~(\ref{D1}) establishes a direct link between the diffusion and local structure, by which one can predict $D$ based on the geometric parameter $k$.

The measured $D$ satisfies Eq.~\ref{D1} well in Fig.~\ref{fig5}(d), indicating that the anisotropy parameter $k$ can provide a single-particle form of the free volume model. To remove the influence of long-range Mermin–Wagner fluctuations in 2D \cite{illing17} in the measurement of $D$, we use the local coordinate for each particle’s position: Particle $i$’s position $\widetilde {\mathbf{r}}_i(t)=\mathbf{r}_i(t)-\sum_j\mathbf{r}_j(t)/N_i$ is relative to the mean position of its $N_i$ nearest neighbors \cite{mazoyer10,vivek17}. The cage-relative mean-square displacement (CR-MSD) is defined as $\Delta r^2_\textrm{CR}(t)=\langle [\widetilde{\mathbf{r}}_i(t) - \widetilde {\mathbf{r}}_i(0) ]^2 \rangle$ \cite{mazoyer10,vivek17},  where $\langle \quad\rangle$ is the ensemble average. The long-time $D$ is measured from the slope of the CR-MSD at $t>10^4$ [Supplemental Fig.~S5(b)]. 

\section{Discussion}
In this work, we propose a nonagnostic structure parameter $k$ with a simple geometrical meaning to characterize the cage anisotropy. $\bar{k}$ exhibits different behavior in liquid and glass regimes, and the structural anisotropic–isotropic transition coincides with the dynamic MCT glass transition at $\phi_\mathrm{c}$. This is robust in both 2D and 3D systems [Figs.~\ref{fig1}(c) and \ref{fig1}(d) and Supplementary Fig.~S1], providing a structural signature at the MCT glass transition point. Moreover, we find an MCT-like relation Eq.~(\ref{mct}) that links the dynamics to local structures at the single-particle level [Fig.~\ref{fig5}(c)]. 

Two applications of $k$ are further studied: (1) The structural power law in glasses \cite{wang09,zeng14,chen15,zeng16,xia17,zhang20} is generalized to the liquid regime with $d>D$ (Fig.~\ref{fig3}). It is qualitatively explained by the free volume change associated with $k$. (2) Substituting $k$ into the classical free-volume model yields Eq.~(\ref{D1}), which can explain and predict the diffusion constant of $D$ from the structure at the particle scale [Eq.~(\ref{D1}) and Fig.~\ref{fig5}(d)]. $k$ reflects the local free volume. $\mathbf{R}_\mathrm{max}$ reflect a particle’s dynamical propensity, that is, particles tend to move along $\pm\mathbf{R}_\mathrm{max}$ because more free volumes exist along this direction. Thus, $k$ and $\mathbf{R}_\mathrm{max}$ can connect to mechanical behaviors of glass, such as soft spot \cite{manning11,Cubuk15} and shear transformation zone \cite{schall07,Richard21}. Moreover, we expect that $k$ is also applicable in characterizing the crystallization \cite{gasser01,zaccarelli09,zhang14,gao21} and melting transitions \cite{wang12melting}, in which the local cages will change from anisotropic in liquids to isotropic in crystals.

\section{ACKNOWLEDGEMENTS}
This work was supported by the National Natural Science Foundation of China (No. 12274336), the Key R\&D Project of Shaanxi Province (No. 2022GY-400), the Fundamental Research Funds for Central Universities (No. xxj032021001), Hong Kong Research Grants Council (No. CRF-C6016-20G and C6021-19EF), and the Guangdong Basic and Applied Basic Research Foundation (No. 2020B1515120067).

\bibliographystyle{unsrt}  
%\bibliography{references}  %%% Remove comment to use the external .bib file (using bibtex).
%%% and comment out the ``thebibliography'' section.
%\bibliography{draft}

\end{document}